\documentclass[a4paper,10pt,preprint,5p]{elsarticle}
\usepackage{hyperref}
\usepackage[utf8]{inputenc}
\usepackage{amsmath}
\usepackage{amsfonts}
\usepackage{amssymb}
\usepackage{graphicx}
\usepackage{color}
\usepackage{url}
\usepackage{tablefootnote}
\usepackage{caption}
\usepackage{marvosym}
\usepackage{subcaption}
\usepackage{algorithm}
\usepackage{float}
\usepackage{pdflscape}

\usepackage{balance}
\usepackage{listings}
\usepackage{fancybox}
\usepackage{multirow}
\usepackage{cleveref}

\usepackage{amsmath,color,soulutf8,longtable,colortbl,setspace,ifthen,xspace,url,pdflscape, amsmath}

\journal{the Journal of Systems and Software}

\newcommand{\ie}{\textit{i.e.,}}
\newcommand{\eg}{\textit{e.g.,}}

\begin{document}

\begin{frontmatter}

\title{On Testing Machine Learning Programs}

\author[swat]{Houssem Ben Braiek}
\ead{houssem.ben-braiek@polymtl.ca}

\author[swat]{Foutse Khomh}
\ead{foutse.khomh@polymtl.ca}

\address[swat]{SWAT Lab., Polytechnique Montr\'{e}al, Canada}

\begin{abstract}
Nowadays, we are witnessing a wide adoption of Machine learning (ML) models in many safety-critical systems, thanks to recent breakthroughs in deep learning and reinforcement learning. Many people are now interacting with systems based on ML every day, \emph{e.g.,} voice recognition systems used by virtual personal assistants like Amazon Alexa or Google Home. As the field of ML continues to grow, we are likely to witness transformative advances in a wide range of areas, from finance, energy, to health and transportation. Given this growing importance of ML-based systems in our daily live, it is becoming utterly important to ensure their reliability. Recently, software researchers have started adapting concepts from the software testing domain (\emph{e.g.,} code coverage, mutation testing, or property-based testing) to help ML engineers detect and correct faults in ML programs. This paper reviews current existing testing practices for ML programs. First, we identify and explain challenges that should be addressed when testing ML programs. Next, we report existing solutions found in the literature for testing ML programs. Finally, we identify gaps in the literature related to the testing of ML programs and make recommendations of future research directions for the scientific community. We hope that this comprehensive review of software testing practices will help ML engineers identify the right approach to improve the reliability of their ML-based systems. We also hope that the research community will act on our proposed research directions to advance the state of the art of testing for ML programs.
\end{abstract}

\begin{keyword}
Machine Learning \sep Data Cleaning \sep Feature Engineering Testing\sep Model Testing \sep Implementation Testing
\end{keyword}

\end{frontmatter}

\section{Introduction}
\begin{figure*}[t]
\centering
\captionsetup{justification=centering}
\includegraphics[scale=0.75]{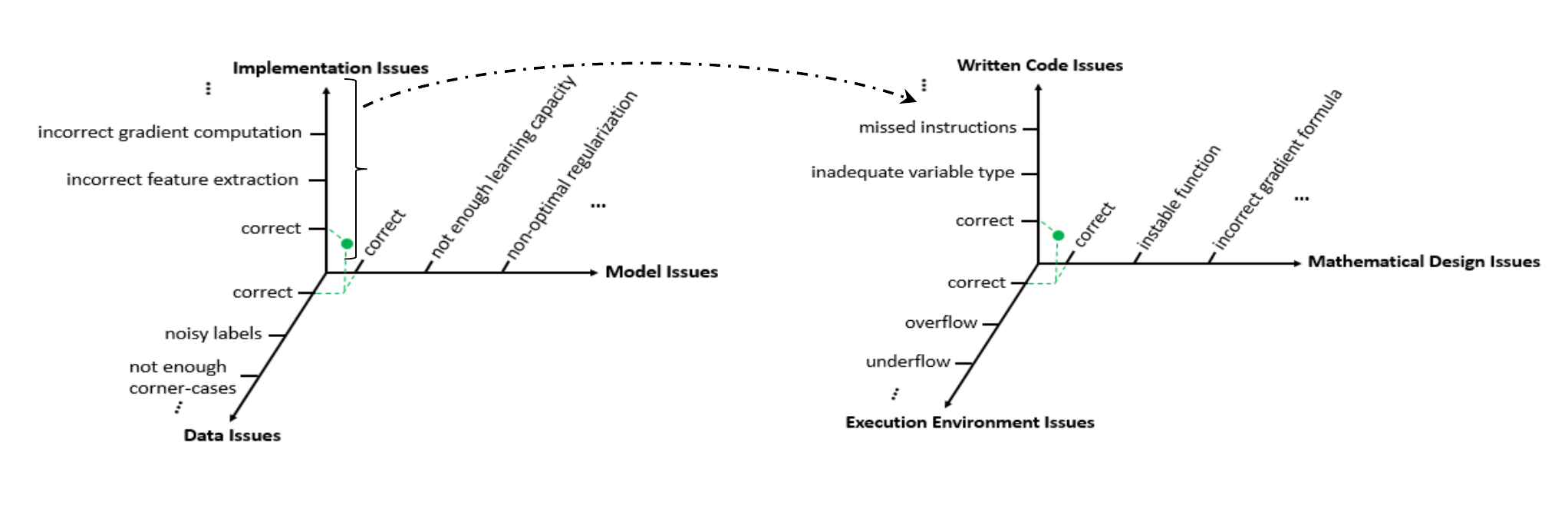}
\caption{Testing Space of ML programs}
\label{fig:testing_space}
\end{figure*}

Machine learning (ML) is increasingly deployed in large-scale and critical systems thanks to recent breakthroughs in deep learning and reinforcement learning. We are now using software applications powered by ML in critical aspects of our daily lives; from finance, energy, to health and transportation. However, detecting and correcting faults in ML programs is still very challenging as evidenced by the recent Uber's car incident that resulted in the death of a pedestrian\footnote{\url{https://www.theguardian.com/technology/2018/may/08/ubers-self-driving-car-saw-the-pedestrian-but-didnt-swerve-report}}. The main reason behind the difficulty to test ML programs is the shift in the development paradigm induced by ML and AI. Traditionally, software systems are constructed deductively, by writing down the rules that govern the behavior of the system as program code. However, with ML, these rules are inferred from training data (\ie{}, they are generated inductively). This paradigm shift in application development makes it difficult to reason about the behavior of software systems with ML components, resulting in systems that are intrinsically challenging to test and verify, given that they do not have (complete) specifications or even source code corresponding to some of their critical behaviors. In fact some ML programs rely on proprietary third-party libraries like Intel Math Kernel Library for many critical operations. A defect in a ML  program may come from its training data, program code, execution environment, or third-party frameworks. Compared with traditional software, the dimension and potential testing space of a ML programs is much more larger as shown in Figure~\ref{fig:testing_space}. Current existing software development techniques must be revisited and adapted to this new reality.

In this paper, we survey existing testing practices that have been proposed for ML programs, explaining the context in which they can be applied and their expected outcome. We also, identify gaps in the literature related to the testing of ML programs and suggest future research directions for the scientific community. This paper makes the following contributions:

\begin{itemize}
\item We present and explain challenges related to the testing of ML programs that use differentiable models.
\item We provide a comprehensive review of current software testing practices for ML programs.
\item We identify gaps in the literature related to the testing of ML programs and provide future research directions for the scientific community.
\end{itemize}

To the best of our knowledge, this is the most comprehensive study of testing practices for ML programs.
\textbf{The rest of the paper is organized as follows.} Section~\ref{sec:background} provides background knowledge on the ML programs. Section~\ref{sec:design} presents research trends in ML application testing. Finally, Section~\ref{sec:conclusion} concludes the paper.


\section{Background on Machine Learning Model}\label{sec:background}
In this section, we provide background information about Machine Learning (ML) programs and explain challenges that should be addressed when testing ML programs. \\
The first step when constructing a ML component is to collect data from which concepts and hidden patterns can be learned using some algorithms. Most machine learning algorithms require huge volume of data to be able to converge and make meaningful inferences, which makes data collection a challenging step. Once data is collected, often it has to be pre-processed before it can be used for learning. Failing to complete this pre-processing properly is likely to result in noisy data which can significantly affect the quality of trained models. After this pre-processing step, important features are identified from the data. These features also often need to be processed before they can be used in a learning algorithm. Inadequate feature engineering (\ie{} failure to process the features adequately) is also likely to result in poor models. Once, the data is cleaned, and features are extracted properly, a learning algorithm is used to infer relations capturing hidden patterns in the data. During this learning process, the parameters of the algorithm are tuned to fit the input data through an iterative process, during which the performance of the model is assessed continuously. A poor choice of parameter or an ineffective model testing mechanism will also result in a poor model. After training and testing steps, the model is deployed in a production environment which can be different from the training and testing environments. In production, the model often has to interact with the other components of the application. 
There are two main source of faults in ML programs : the data and the model. For each of these two dimensions (\ie{} data and model), there can be errors both at the conceptual and implementation levels, which makes fault location very challenging. Approaches that rely on tweaking variables and watching signals from execution are generally ineffective because of the exponential increase of the number of potential faults locations. In fact, if there are $n$ potential faults related to data, and $m$ potential faults related to the model, we have $n \times m$ total possible faults locations when data is feed in the model. This number grows further when the model is implemented as code, since more types of faults can be introduced in the program at that stage as illustrated on Figure~\ref{fig:testing_space}. Systematic and efficient testing practices are necessary to help ML engineers detect and correct faults in ML programs. In the following we present potential errors that can occur in data and models, both at design and implementation levels.

\subsection{Data Engineering: Challenges and Issues}\label{data_construction}
Data is an essential artifact when building ML models. It often comes from a variety of sources \eg{} mainframe databases, sensors, IOT devices, and software systems, and is presented in different formats (\eg{} various media types). It can be structured (such as database records) or unstructured (such as raw texts) and is delivered to ML models either in batch (\eg{} discrete chunks from mainframe databases and file systems) and/or real-time (\eg{} continuous flow from IOT devices or Stream REST API). ML engineers generally have to leverage complementary automated tools that support batch and/or real-time data ingestion strategies, to collect data needed for training ML models. \paragraph{\textbf{Conceptual issues}} Once data is gathered, cleaning data tasks are required to ensure that the data is consistent, free from redundancy and given a reliable starting point for statistical learning. Common cleaning tasks include: (1) removing
invalid or undefined values (\ie{} Not-a-Number, Not-Available), duplicate rows, and outliers that seems to be too different from the mean value); and (2) unifying the variables' representations to avoid multiple data formats and mixed numerical scales. This can be done by data transformations such as normalization, min-max scaling, and data format conversion. This pre-processing step allows to ensure a high quality of raw data, which is very important because decisions made on the basis of noisy data could be wrong. In fact, recent sophisticated ML models endowed by a high learning capacity are highly sensitive to noisy data~\cite{activeclean}. This brittleness makes model training unreliable in the presence of noisy data, which often results in poor prediction performances~\cite{DirtyImpacts}.

Once data is cleaned, pertinent features that describe the structures inherent in the data entities are selected from a large set of initial variables in order to eliminate redundant or irrelevant ones. This selection is based on statistical techniques such as correlation measures and variance analysis. Afterwards, these features are encoded to particular data structures to allow feeding them to a chosen ML model that can handle the features and recognize their hidden related patterns in an effective way. The identified patterns represent the core logic of the model. Changes in data (\ie{} the input signals) are likely to have a direct impact on these patterns and hence on the behavior of the model and its corresponding predictions. Because of this strong dependence on data, ML models are considered to be data-sensitive or data-dependent algorithms. A poor selection of features can impact a ML system negatively. Sculley et al.~\cite{HiddenMLtechDebt} report that unnecessary dependencies to features that contribute with little or no value to the model quality can generate vulnerabilities and noises in a ML system. Examples of such features are : \textit{Epsilon Feature}, which are features that have no significant contribution to the performance of the model, \textit{Legacy Feature}, which are features that lost their added information value on model accuracy improvement when other more rich features are included in the model, or \textit{Bundled Features}, which are groups of features that are integrated to a ML system simultaneously without a proper testing of the contribution of each individual feature.
\paragraph{\textbf{Implementation issues}} To process data as described above, ML engineers implement data pipelines containing components for data transformations, validation, enrichment, summarization, and--or any other necessary treatment. Each pipeline component is separated from the others, and takes in a defined input, and returns a defined output that will be served as input data to the next component in the pipeline. Data pipelines are very useful in the training phase as they help process huge volume of data. They also transform raw data into sets of features ready-to-be consumed by the models. Like any other software component, data pipelines are not immunized to faults. There can be errors in the code written to implement a data pipeline. Sculley et al.\cite{MLtechDebt} identified two common problems affecting data pipelines:
\begin{itemize}
\item \textit{Pipelines jungles} which are overly convoluted and unstructured data preparation pipelines. This appears when data is transformed from multiple sources at various points through scraping, table joins and other methods without a clear, holistic view of what is going on. Such implementation is prone to faults since developers lacking a good understanding of the code are likely to make mistakes. Also, debugging errors in such code is challenging.
\item \textit{Dead experimental code paths} which happens when code is written for rapid prototyping to gain quick turnaround times by performing additional experiments simply by tweaks and experimental code paths within the main production code. The remnants of alternative methods that have not been pruned from the code base could be executed in certain real world situations and create unexpected results.
\end{itemize}

For systems that rely on mini-batch stochastic optimizers to estimate the model's parameters, like deep learning models, another data-related component is required in addition to data pipelines, \ie{} the Data Loader. This component is responsible of the generation of the batches that are used to assess the quality of samples, during the training phase. It is also prone to errors. 


\subsection{Model Engineering: Challenges and Issues}\label{model_construction}
Once ML engineers have collected and processed the data, they proceed to finding the appropriate statistical learning model that could fit the available data in order to build its own logic and solve the given problem.
A wide range of statistical models can be acquired and--or extended to suit different classification and regression purposes. There are simple models that make initial assumptions about hidden relationships in the data. For example, the linear regression assumes that the output can be specified as a linear combination of features and the SVM classifier assumes that there is an hyperplane with maximum margin that can optimally separate the two classes of the output. Besides, there are more complex models such as neural network models, which usually do not make assumptions about the relationship between incoming pairs of data. In fact, Neural Network is structured in terms of interconnected layers of computation units, much like the neurons and their connectivity in the brain. Each neuron includes an activation function (\ie{} a non-linear transformation) to a weighted sum of input values with bias. The predicted output is calculated by computing the outputs of each neuron through the network layers in a feed-forward manner. At the end, Neural Network's mapping function can be seen as a composite function encoding a sequence of linear transformations and their following non-linear ones. The strength of this complex model lies in the universal approximation theorem. A feed-forward network with a single hidden layer containing a finite number of neurons can approximate any continuous function, under mild assumptions on the activation function. However, this does not indicate how much neurons are required and the number can evaluate exponentially with respect to the complexity of formulating the relationships between inputs and outputs. Deep neural networks, which are the backbone behind deep learning, use a cascade of multiple hidden layers to reduce the number of neurons required in each layer. 

Regardless of the model, ML programs discover hidden patterns in training data and build a mathematical model that makes predictions or identifications on future unseen data, using these patterns. The learning aspect resides in the model fitting which is an iterative process during which little adjustments are made repeatedly, with the aim of refining the model until it predicts mostly the right outputs. Generally, supervised machine learning algorithms are based on a differentiable model that trains itself on input data through optimization routines using gradient learning, to create a better model or probably the best fitted one (\ie{} this could happen with a convex objective function or advanced update steps). The principal components of a differentiable model are:
\begin{itemize}
\item \textbf{Parameters}, on which the model depends to make its internal calculation and to provide its prediction outputs. Therefore, these signals or factors are inner variables of the model that the ML program gradually adjusts on its own through successive training iterations to build its logic and form its decision about future data. For example, weights and biases used by simple linear regression models or by Neural Network's neurons are parameters.
\item \textbf{Loss Function}, at its core, represents a mathematical function given a real value that provide an estimation on how the model is performing in terms of learning goals. It assesses the sum or the average of the distance measure between predicted outputs and actual outcomes. As indicated by its name, it reflects the cost or the penalty of making bad predictions. If the model's predictions are perfect, the loss is zero; otherwise, the loss is greater.
\item \textbf{Regularization}, assembles techniques, that penalize the model's complexity to prevent overfitting. An example of regularization technique is the use of $L2$ norm in linear regression models, which keeps smaller overall weight values, relying on a prior belief that weights should be small and normally distributed around zero. Another example of regularization technique is the \emph{dropout} in neural networks, which allows removing a random selection of a percentage of neurons from training during an iteration. This is to prevent models with high learning capacity from memorizing the peculiarities of the training data, which would result in complex models that overfit the training data. 
    To guarantee a model's generalization ability, two parallel objectives should be reached: (1) build the best-fitted model \ie{} lowest loss and (2) keep the model as simple as possible \ie{} strong regularization.
\item \textbf{Optimizer}, adjusts iteratively the internal parameters of the model in a way that reduces the objective function, \ie{} includes generally the loss function plus a regularisation term. The most used optimizers are based on gradient descent algorithms, which minimize gradually the objective function by computing the gradients of loss with respect to the model's parameters and updates their values in direction opposite to the gradients until finding a local or the global minimum. The objective function has to be globally continuous and differentiable. It is desirable that this function be also strictly convex and has exactly one local minimum point, which is also the global minimum point. A great deal of research in ML has focused on formulating various problems as convex optimization problems and solving them efficiently. Deep neural networks are never convex functions but they are very successful because many variations of gradient descent have a high probability of finding reasonably good solutions anyway, even though these solutions are not guaranteed to be global minimums.
\item \textbf{Hyperparameters}, represent the model's parameters that are constant during the training phase and which can be fixed before running the fitting process. By learning a ML model, the ML engineers identify a specific point in the model space with some desirable behavior. In fact, choosing in prior the model's hyperparameters, such as the number of layers and neurons in each layer for neural network, the learning rate for gradient descent optimizer or the regularization rate, allows to identify a subset of the model space to search. It lasts to use the optimizer with aim of finding the best fitted model from the selected subset through parameters adjustments. Similarly, hyperparameters tuning is highly recommended to find the composition of hyper-parameters that helps the optimization process finding the best-fitted model. The most used search methods are (1) \emph{Grid search}, which explores all possible combinations from a discrete set of values for each hyperparameter; (2) \emph{Random search} which samples random values from a statistical distribution for each hyperparameter, and (3) \emph{Bayesian optimization} which chooses iteratively the optimal values according to the posterior expectation of the hyperparameter. This expectation is computed based on previously evaluated values, until converging to an optimum.
 \end{itemize}

To train a differentiable model one needs three different datasets : \emph{training dataset}, \emph{validation dataset}, and \emph{testing dataset}. After readying these data sets, ML engineers set initial hyperparameters values and select loss functions, regularization terms, and gradient-based optimizers following best practices or guidelines from other works that addressed a similar problem. Training a ML model using an optimizer consists in gradually minimizing the loss measure plus the regularization term with respect to the training dataset. Once the model's parameters are estimated, hyperparameters are tuned by evaluating the model performance on the \emph{validation dataset}, and selecting the next hyperparameters values according to a search-based approach that aims to optimize the performance of the model. This process is repeated using the newly selected hyperparameters until a best-fitted model is obtained. This best-fitted model is therefore tested using the testing dataset (which should be different from training and validation datasets). 
\paragraph{\textbf{Conceptual issues}} One key assumption behind the training process of supervised ML models is that the \emph{training dataset}, the \emph{validation dataset}, and  the \emph{testing dataset}, which are sampled from manually labeled data, are representative samples of the underlying problem. Following the concept of Empirical Risk Minimization (ERM), the optimizer allows finding the fitted model that minimizes the empirical risk; which is the loss computed over the training data assuming that it is a representative sample of the target distribution. The empirical risk can correctly approximates the \emph{true risk} only if the training data distribution is a good approximation of the true data distribution (which is often out of reach in real-world scenarios). 
The size of the training dataset has an impact on the approximation goodness of the true risk, \ie{} the larger is a training data, the better this approximation will be. However, manual labeling of data is very expensive, time-consuming and error-prone. Training data sets that deviate from the reality induce erroneous models.
\paragraph{\textbf{Implementation issues}}
ML algorithms are often proposed in pseudo-code formats that include jointly scientific formula and algorithmic rules and concepts. When it comes to implementing ML algorithms, ML engineers sometimes have difficulties understanding these formulas, rules, or concepts. Moreover, because there is no `test oracle'' to verify the correctness of the estimated parameters (\ie{} the computation results) of a ML model, it is difficult to detect faults in the learning code. Also, ML algorithms often require sophisticated numerical computations that can be difficult to implement on recent hardware architectures that offer high-throughput computing power. Weyuker~\cite{nonTestableProgs} identified three distinct sources of errors in scientific software programs such as ML programs. 
\begin{enumerate}
\item \textbf{The mathematical model used to describe the problem}\\
A ML program is a software implementation of a statistical learning algorithms that requires substantial expertise in mathematics (\eg{} linear algebra, statistics, multivariate analysis, measure theory, differential geometry, topology) to understand its internal functions and to apprehend what each component is supposed to do and how we can verify that they do it correctly. Non-convex objectives can cause unfavorable optimization landscape and inadequate search strategies. Model mis-specifications or a poor choice of mathematical functions can lead to undesired behaviors. For example, many ML algorithms require mathematical optimizations that involve extensive algebraic derivations to put mathematical expressions in closed-form. This is often done through informal algebra by hand, to derive the objective or loss function and obtain the gradient. Generally, we aim to adjust model parameters following the direction that minimizes the loss function, which is calculated based on a comparison between the algorithmic outputs and the actual answers. Like any other informal tasks, this task is subject to human errors. The detection of these errors can be very challenging, in particular when randomness is involved. 
Stochastic implementation errors can persist indefinitely without detection, since some errors may be masked by the distributions of random variables and may require writing customized statistical tests for their detection.\\
To avoid overfitting, ML engineers often add a regularization loss (\eg{} norm $L_2$ penalty on weights). This regularization term which has a simple gradient expression can overwhelm the overall loss; resulting in a gradient that is primarily coming from it. Which can make the detection of errors in the real gradient very challenging. 
Also, non-deterministic regularization techniques, such as dropout in neural network can cause high-variance in gradient values and further complicate the detection of errors, especially when techniques like numerical estimation are used.\\
\item \textbf{The program written to implement the computation}\\
The program written to implement a mathematical operation can differ significantly from the intended mathematical semantic when complex optimization mechanisms are used. 
Nowadays, most ML programs leverage rich data structures (\eg{} data frames) and high performance computing power to process massive data with huge dimensionality. The optimization mechanisms of these ML programs is often solved or approximated by linear methods that consists of regular linear algebra operations involving for example multiplication and addition operations on vectors and matrices. This choice is guided by the fact that high performance computers can leverage parallelization mechanisms to accelerate the execution of programs written using linear algebra. However, to leverage this parallelism on Graphics Processing Unit (GPU) platforms for example, one has to move to higher levels of abstraction. The most common abstractions used in this case are tensors, which are multidimensional arrays with some related operations. Most ML algorithms nowadays are formulated in terms of matrix-vector, matrix-matrix operations, and tensor-based operations (to extract a maximum of performance from the hardware). 
Also, ML models are more and more sophisticated, with multiple layers containing huge number of parameters each. For such models, the gradient which represents partial derivatives that specify how the loss function is altered through individual changes in each parameter is computed by grouping the partial derivatives together in multidimensional data structures such as tensors, to allow for more straight forward optimized and parallelized calculations. This large gap between the mechanics of the high performance implementation and the mathematical model of one ML algorithm makes the translation from scientific pseudo-code to highly-optimized program difficult and error-prone. It is important to test that the code representations of these algorithms reflect the algorithms accurately. 
\item \textbf{Features of the environment such as round-off error}\\
The computation of continuous functions such as gradient on discrete computational environments like a digital computer incur some approximation errors when one needs to represent infinitely many real numbers with a finite number of bit patterns. These numerical errors of real numbers' discrete representations can be either overflow or underflow. An overflow occurs when numbers with large magnitude are approximated as $+\infty$ or $-\infty$, which become not-a-number values if they are used for many arithmetic operations. An underflow occurs when numbers near zero are rounded to zero. This can cause the numerical instability of functions such as division (\ie{} division by a zero returns not-a-number value) or logarithm (\ie{} logarithm of zero returns $-\infty$ that could be transform into not-a-number by further arithmetic). \\
Hence, it is not sufficient to validate a scientific computing algorithm theoretically since rounding errors can lead to the failure of its implementation. Rounding errors on different execution platforms can cause instabilities in the execution of ML models if their robustness to such errors is not handled properly. Testing for these rounding errors can help select adequate mathematical formulations that minimize their effect.
\end{enumerate}


When implementing ML programs, developers often rely on third-party libraries for many critical operations. These libraries provide optimized implementations of highly intensive computation functions, allowing ML developers to leverage distributed infrastructures such as high-performance computing (HPC) clusters, parallelized kernels executing on high-powered graphics processing units (GPUs), or to merge these technologies to breed new clusters of multiple GPUs. However, a misuse of these libraries can result in faults that are hard to detect. For example, a copy-paste of a routine that creates a neural network layer, without changing the corresponding parameters (\ie{} weights and biases variables) would result in a network where the same parameters are shared between two different layers, which is problematic. Moreover, this error can remain in the code unnoticed for a long time. Which is why it is utterly important to test that configuration choices do not cause faults or instabilities in ML programs. In the following we explain the three main categories of libraries that exist today and discuss of their potential misuse.

\begin{itemize}
\item \textbf{High-level ML libraries:} A high-level ML library emphasizes the ease of use through features such as state-of-the-art supervised learning algorithms, unsupervised learning algorithms, and evaluation methods (\eg{} confusion matrix and cross-validation). It serves as an interface and provides a high level of abstraction, where ML algorithms can be easily configured without hard-coding. Using such library, ML developers can focus on converting a business objective into a ML-solvable problem. However, ML developers still need to test the quality of data and ensure that it is conform to the requirements of these built-in functions, such as input data formats and types. Moreover, a poor configuration of these provided algorithms could result in unstable or misconceived models. For example, choosing a sigmoid as activation functions in a deep neural network can cause the saturation of neurons and consequently, slowing down the learning process.  Therefore, after finishing the configuration, developers need to set up monitoring routines to watch internal variables and metrics during the execution of the provided algorithms, in order to check for possible numerical instabilities, or suspicious outputs.
\item \textbf{Medium-level ML libraries:} Medium-level libraries provide machine learning or deep learning routines as ready-for-use features, such as numerical optimization techniques, mathematical function and automatic differentiation capabilities, allowing ML developers to not only configure the pre-defined ML algorithms, but also to use the provided routines to define the flow of execution of the algorithms. This flexibility allows for easy extensions of the ML models using more optimized implementations. However, this ability to design the algorithm and its computation flow through programming variables and native loops increases the risk of faults and poor coding practices, as it is the case for any traditional program. 
\item \textbf{Low-level ML libraries:} Contrary to high level and medium level libraries, this family of libraries do not provide any pre-defined ML feature, instead, they provide low level computations primitives that are optimized for different platforms. They offer powerful $N$-dimensional arrays, which are commonly used in numerical operations and linear algebra routines for a high level of efficiency. They also help with data processing operations such as slicing and indexing. ML developers can use these libraries to build a new ML algorithm from scratch or highly-optimized implementations of particular algorithms for specific contexts or hardwares. However, this total control on the implementation is not without cost. Effective quality assurance operations such as testing and code reviews are required to ensure bug free implementations. ML developers need strong backgrounds in mathematics and programming to be able to work efficiently with these low level libraries.
\end{itemize}

The amount of code that is written when implementing a ML program depends on the type of ML library used. The more a developer uses high-level features from libraries, the more he has to write glue code to integrate incompatible components, which are putted together into a single implementation. Sculley et al.~\cite{MLtechDebt} observed that the amount of this glue can account for up to 95\% of the code of certain ML programs. This code should be tested thoroughly to avoid faults.


\section{Research Trends in ML Application Testing}\label{sec:design}
In this section, we analyze research works that proposed testing techniques for ML programs. We organize the testing techniques 
in two categories based on the intention behind the techniques, \ie{} techniques that aim to detect conceptual and implementation errors in data, and techniques that focus on ensuring correct conception and implementation of ML models. In each of these categories, we divide the techniques in sub-groups based on the concepts used in the techniques. 
Next, we discuss the fundamental concepts behind each proposed techniques, explaining the types of errors that can be identified using them while also outlining their limitations.

\subsection{Approaches that aim to detect conceptual and implementation errors in data}
The approaches proposed in the literature to test the quality of data addresses both conceptual and implementation issues, therefore, we discuss these two aspects together in this section. 
The most common technique used to test the quality of data is the analysis-driven data cleaning that consists of applying analytical queries (\eg{} aggregates, advanced statistical analytic, etc.) in order to detect errors and perform adequate transformations.\\
In this approach, aggregations such as sum or count and central tendencies such as mean, median or mode are used to verify if each feature's distribution matches expectation. For example, one can check that features take on their usual set or range of values, and the frequencies of these values in the data.\\
After this initial verification, advanced statistical analyses such as hypothesis testing and correlation analysis are applied to verify correlations between pair of features and to assess the contribution of each feature in the prediction or explanation of the target variable. 
The benefit of each feature can also be estimated by computing the proportion of its explained variance with respect to the target output or by assessing the resulting accuracy of the model when removing it in prior to the fitting process. Besides, when assessing the contribution of each feature to the model, it is recommended to take into account the added inference latency and RAM usage, more upstream data dependencies, and additional expected instability incurred by relying on that feature. It is important to consider whether this cost is worth paying when traded off against the provided improvement in model quality.

Recent research work by Krishnan et al.~\cite{sampleclean} remarked that these aggregated queries can sometimes diminish the benefits of data cleaning. They observed that cleaning small samples of data often suffices to estimate results with high accuracy. They also observed that the power of statistical anomaly detection techniques rapidly deteriorates in the high-dimensional feature-spaces.\\
This comes from the fact that the aforementioned analysis-driven data cleaning operations require data queries in order to calculate the aggregate values and correlation measures. Indeed, performing many data queries and cleaning operation on the entire dataset could be impractical with huge amount of training datasets that likely contain dirty records. Moreover, ML developers often face difficulties to establish the data cleaning process. 
To address these issues, Krishnan et al. proposed ActiveClean~\cite{activeclean}, an interactive data-cleaning framework for ML models that allows ML developers to improve the performance of their model progressively as they clean the data. The framework has an embedding process that firstly samples a subset of likely dirty records from training data using a set of optimizations, which includes importance weighting and dirty data detection. Secondly, the ML developer is interactively invited to transform or remove each data instance from the selected batches of probably dirty data. Finally, the framework updates the model's parameters and continues the training using partially cleaned data. This process is repeated until no potential dirty instances could be detected. 
With ActiveClean, developers are still responsible for defining data cleaning operations. The framework only decides where to apply these operations. 
Recently, Krishnan et al.~\cite{boostclean} proposed a full-automated framework, BoostClean, to establish a pipeline of data transformations that allow cleaning efficiently the data in order to fit well the model. BoostClean finds automatically the best combination of dirty data detection and repair operations by leveraging the available clean test labels in order to improve model accuracy. It selects this ensemble from extensible libraries : (1) pre-populated general detection functions, allow identifying numerical outliers, invalid and missing values, checking whether a variable values match the column type signature, and detecting effectively text errors in string-valued and categorical attributes using word embedding; (2) a pre-populated set of simple repair functions that can be applied to records identified by a detector's predicate, such as impute a cell value with one central tendency (\ie{} mean, median and mode value), or discard a dirty record from the dataset. Thus, the boosting technique, which combines a set of weak learners and estimates their corresponding weights to spawn a single prediction model, is applied to solve the problem of detecting the optimal sequence of repairs that could best improve the ML model by formulating it as an ensembling problem. It consists of generating a new model trained on input data with new additional cleaned features and selecting the best collection of models that collectively estimate the predict label.
Krishnan et al. evaluated their proposed framework on 8 ML datasets from Kaggle and the UCI repository which contained real data errors. They showed that BoostClean can increase the absolute prediction accuracy by 8-9\% over the best non-ensemble alternatives, including statistical anomaly detection and constraint-based techniques.\\
By automating the selection of cleaning operations, BoostClean significantly simplifies the data cleaning process, however, this framework is resource consuming as it requires the use of multiple models and boosting techniques. Moreover, the evaluation of the embedded cleaning process results on new datasets is challenging because the creation of the cleaning data pipeline is driven by pure statistical analysis. \\
Hynes et al.\cite{datalinter} inspired by code linters, which are well-known software quality tools, introduced data linter to help ML developers track and fix issues in relation to data cleaning, data transformation and feature extraction. The data linter helps reduce the human burden by automatically generating issues explanations and building more sophisticated human-interactive loop processing. First, it inspects errors in training datasets such as scale differences in numerical features, missing or illegal values (\eg{} NaN), malformed values of special string types (\eg{} dates), and other problematic issues or inefficiencies discussed in Section~\ref{data_construction}. The inspection relies on data's summary statistics, individual items inspection, and column names given to the features. Second, given the detected errors and non-optimal data representations, it produces a warning, a recommendation for how to transform the feature into a correct or optimal feature, and a concrete instance of the lint taken directly from the data. Data linter guides its users in their cleaning data and features engineering process through providing actionable instructions of how individual features can be transformed to increase the likelihood that the model can learn from them. The main strength of this tool resides in the semi-automated data engineering process and the fact that it can be applied to all statistical learning models and several different data types. The proposed data linter has the ability to infer semantic meaning/intent of a feature as a complement to statistical analysis with the aim of providing specific and comprehensible feature engineering recommendations to ML developers.\\
As mentioned in Section~\ref{data_construction}, a Data loader is often required for systems that rely on mini-batch stochastic optimizers to estimate the model's parameters. To test the reliability of this component, the following best practices are often used:
(1) Shuffling of the dataset before starting the batches generation. This action is recommended to prevent the occurrence of one single label batch (\ie{} sample of data labeled by the same class) which would negatively affect the efficiency of mini-batch stochastic optimizers in finding the optimal solution. In fact, a straightforward extraction of batches in sequence from data ordered by label or following a particular semantic order, can cause the occurrence of one single label batch. (2) Checking the predictor/predict inputs matching.
A random set of few inputs should be checked to verify if they are correctly connected to their labels following the shuffle of data; (3) reduce class imbalance. This step is important to  
keep the class proportions relatively conform to the totality of training data.

\subsection{Approaches that aim to detect conceptual and implementation errors in ML models}
As discussed in Section~\ref{model_construction} and illustrated on Figure~\ref{fig:testing_space}, errors in ML models can be due to conceptual mistakes when creating the model or implementation errors when writing the code corresponding to the model. In the following, we discuss testing approaches that focus on these two aspects, separately.
\subsubsection{\textbf{Approaches that aim to detect conceptual errors in ML models}}
Approaches in this category assume that the models are implemented into programs without errors and focus on providing mechanisms to detect potential errors in the calibration of the models. These approaches can be divided in two groups: black-box and white-box approaches~\cite{blackAndWhite}. Black-box approaches are testing approaches that do not need access to the internal implementation details of the model under test. These approaches focus on ensuring that the model under test predicts the target value with a high accuracy, without caring about its internal learned parameters. White-box testing approaches on the other hand take into account the internal implementation logic of the model. The goal of these approaches is to cover a maximum of specific spots (\eg{} neurons) in models. In the following, we elaborate more on approaches from these two groups.

\paragraph{\textbf{A: Black-box testing approaches for ML models}}
The common denominator to black-box testing approaches is the generation of adversarial data set that is used to test the ML models. These approaches leverage statistical analysis techniques to devise a multidimensional random process that can generate data with the same statistical characteristics as the input data of the model. More specifically, they construct generative models that can fit a probability distribution that best describes the input data. These generative models allows to sample the probability distribution of input data and generate as many data points as needed for testing the ML models. Using the generative models, the input data set is slightly perturbed to generate novel data that retains many of the original data properties. The advantage of this approach is that the synthetic data that is used to test the model is independent from the ML model, but statistically close to its input data. 
Adversarial machine learning is an emerging technique that aims to assess the robustness the machine learning models based on the generation of adversarial examples. The latter are pairs of test inputs that cause a disagreement on their classification labels when close to each other (in terms of a given distance metric).\\
In fact, ML models are designed to identify latent concepts from the training in order to learn how to predict the target value in relation to the unseen test data. However, the fact that both of training and testing data set are assumed to be generated from the same distribution makes the ML system vulnerable with respect to malicious adaptive adversaries that manipulate the input data and violate some of their prior hypotheses. Hence, it is important to test the robustness of ML models to such variations in input data.
Several mechanisms exist for the 
creation of adversarial examples, such as : making small modifications to the input pixels\cite{goodfellow6572explaining}, applying spatial transformations\cite{engstrom2017rotation}, or simple guess-and-check to find misclassified \cite{gilmer2018motivating}.

Recent results \cite{adversarialExps} \cite{deepfool} involving adversarial evasion attacks against deep neural network models have demonstrated the effectiveness of adversarial examples in revealing weaknesses in models. Multiple DNN-based image classifiers that achieved state-of-the-art performance levels on randomly selected dataset where found to perform poorly on synthetic images generated by adding humanly imperceptible perturbations.

One major limitation of these black-box testing techniques is the representativeness of the generated adversarial examples. In fact, many adversarial models that generate synthetic images often apply only tiny, undetectable, and imperceptible perturbations, since any visible change would require manual inspection to ensure the correctness of the model's decision. This can result in strange aberrations or simplified representations in synthetic datasets, which in turn can have a hidden knock-on effects on the performance of a ML model when unleashed in a real-world setting. These black-box testing techniques that rely only on adversarial data (ignoring the internal implementation details of the models under test) often fail to uncover different erroneous behaviors of the model, even after performing a large number of tests. This is because the generated adversarial data often fail to cover the possible behaviors of the model adequately. An outcome that is not surprising given that the adversarial data are generated without considering information about the structure of the models. To help improve over these limitations, ML researchers have developed the white-box techniques described below, which use internal structure specificities to guide the generation of more relevant test cases.

\paragraph{\textbf{B: White-box testing approaches for ML models}}

Pei et al. proposed DeepXplore~\cite{deepxplore}, the first white-box approach for systematically testing deep learning models. DeepXplore is capable of automatically identifying erroneous behaviors in deep learning models without the need of manual labelling. The technique makes use of a new metric named \emph{neuron coverage}, which estimates the amount of neural network's logic explored by a set of inputs. This neuron coverage metric computes the rate of activated neurons in the neural network. It was inspired by the code coverage metrics used for traditional software systems. The approach circumvent the lack of a reference oracle, by using differential testing. Differential testing is a pseudo-oracle testing approach that has been successfully applied to traditional software that do not have a reference test oracle~\cite{differential}. It is based on the intuition that any divergence between programs' behaviors, solving the same problem, on the same input data is a probably due to an error. Therefore, the test process consists of writing at first multiple independent programs to fulfill the same specification. Then, the same inputs are provided to these similar programs, which are considered as cross-referencing oracles. Differences in their executions are inspected to identify potential errors. In a same way, DeepXplore leverages a group of similar deep neural networks that solve the same problem. Perturbations are introduced in inputs data to create many realistic visible differences (\eg{} different lighting, occlusion, etc.) and automatically detect erroneous behaviors of deep neural networks under these circumstances. Applying differential testing to deep learning with the aim of finding a large number of difference-inducing inputs while maximizing neuron coverage can be formulated as a joint optimization problem. DeepXplore performs gradient ascent to solve efficiently this optimization problem using the gradient of the deep neural network with respect to the input. Its objective is to generate test data that provokes a different behavior from the group of similar deep neural networks under test in order to ensure a high neuronal coverage. We noticed that domain-specific constraints are added to generate data that is valid and realistic. In the end of the testing process, the generated data are kept for future training, to have more robustness in the model.

Ma et al.\cite{deepgauge} generalized the concept of \emph{neuron coverage} by proposing DeepGauge, a set of multi-granularity testing criteria for Deep Learning systems. DeepGauge measures the testing quality of test data (whether it being genuine or synthetic) in terms of its capacity to trigger both major function regions as well as the corner-case regions of DNNs(Deep Neural Networks). It separates DNNs testing coverage in two different levels.\\ At the neuron-level, the first criterion is $k$-multisection neuron coverage, where the range of values observed during training sessions for each neuron are divided into $k$ sections to assess the relative frequency of returning a value belonging to each section. In addition, the authors insist on the need for test inputs that are enough different from training data distribution to cover rare neurons' outputs. They introduced the concept of neuron boundary coverage to measure how well the test datasets can push activation values to go above and below a pre-defined bound (\ie{} covering the upper boundary and the lower boundary values). Their design intentions are complementary to Pei et al. in the sense that the $k$-multisection neuron coverage could potentially help to cover the main functionalities provided by DNN. However, the neuron boundary coverage could relatively approximate corner-cases DNN's behaviors.\\
At layer-level, the authors leveraged recent findings that empirically showed the potential usefulness of discovered patterns within the hyperactive neurons, which render relatively larger outputs. On the one hand, each layer allows DNN to characterize and identify particular features from input data and its main function is in large part supported by its top active neurons. Therefore, regarding the effectiveness in discovering issues, test cases should go beyond these identified hyperactive neurons in each layer. On the other hand, DNN provide the predicted output based on pattern recognized from a sequence features, including simple and complex ones. These features are computed by passing the summary information through hidden layers. Thereby, the combinations of top hyperactive neurons from different layers characterize the behaviors of DNN and the functional scenarios covered. Intuitively, test data sets should trigger other patterns of activated neurons in order to discover corner-cases behaviors.\\
In their empirical evaluation, Ma et al. showed that DeepGauge scales well to practical sized DNN models (\eg{} VGG-19, ResNet-50) and that it could capture erroneous behavior introduced by four state-of-the-art adversarial data generation algorithms (\ie{} Fast Gradient Sign Method (FGSM)~\cite{goodfellow6572explaining}, Basic Iterative Method (BIM)~\cite{kurakin2016adversarial}, Jacobian-based Saliency Map Attack (JSMA)\cite{papernot2016limitations} and Carlini/Wagner attack (CW)~\cite{adversarialExps}). Therefore, a higher coverage of their criteria potentially plays a substantial role, in improving the detection of errors in the DNNs. 
These positive results show the possibility to leverage this multi-level coverage criteria to create automated white-box testing frameworks for neural networks.

Building on the pioneer work of Pei et al., Tian et al. proposed DeepTest~\cite{deeptest}, a tool for automated testing of DNN-driven autonomous cars. In DeepTest, Tian et al. expanded the notion of neuron coverage proposed by Pei et al. for CNNs (Convolutional Neural Networks), to other types of neural networks, including RNNs (Recurrent Neural Networks). Moreover, instead of randomly injecting perturbations in input image data, DeepTest focuses on generating realistic synthetic images by applying realistic image transformations like changing brightness, contrast, translation, scaling, horizontal shearing, rotation, blurring, fog effect, and rain effect, etc. They also mimic different real-world phenomena like camera lens distortions, object movements, different weather conditions, etc. They argue that generating inputs that maximize neuron coverage cannot test the robustness of trained DNN unless the inputs are likely to appear in the real-world. They provide a neuron-coverage-guided greedy search technique for efficiently finding sophisticated synthetic tests which capture different realistic image transformations that can increase neuron coverage in a self-driving car DNNs.
To compensate for the lack of a reference oracle, DeepXplore used differential testing. However, DeepTest leverages metamorphic relations (MRs) to create a test oracle that allows it to identify erroneous behaviors without requiring multiple DNNs or manual labeling. Metamorphic testing~\cite{metamorphic} is another pseudo-oracle software testing technique that allows identifying erroneous behaviors by detecting violations of domain-specific metamorphic relations (MR). These MRs are defined across outputs from multiple executions of the test program with different inputs. The application of such test consists of performing some input changes in a certain way that allows testers to predict the output based on identified MRs, so any significant differences in output would break the relation, which would indicate the existence of errors in the program. 
DeepRoad~\cite{deeproad} continued the same line of work as DeepTest, designing a systematic mechanism for the automatic generation of test cases for DNNs used in autonomous driving cars. Data sets capturing complex real-world driving situations is generated and Metamorphic Testing is applied to map each data point into the predicted continuous output. However, DeepRoad differentiates from DeepTest in the approach used to generate new test images. DeepRoad relies on a Generative Adversarial Network (GAN)-based method to provide realistic snowy and rainy scenes, which can hardly be distinguished from original scenes and cannot be generated by DeepTest using simple affine transformations. Zhang et al. argue that DeepTest synthetic image transformations, such as adding blurring/fog/rain effect filters, cannot simulate complex weather conditions. They claim that DeepTest's produced road scenes may be unrealistic, because simply adding a group of lines over the original images cannot reflect the rain condition or mixing the original scene with the scrambled ``smoke'' effect does not simulate the fog. To solve this lack of realism in generated data, DeepRoad leveraged a recent unsupervised DNN-based method (\ie{} UNIT) which is based on GANs and VAEs, to perform image-to-image transformations. UNIT can project images from two different domains (\eg{} a dry driving scene and a snowy driving scene) into a shared latent space, allowing the generative model to derive the artificial image (\eg{} the snowy driving scene) from the original image (\eg{} the dry driving scene). Evaluation results show that the generative model used by DeepRoad successfully generates realistic scenes, allowing for the detection of thousands of behavioral inconsistencies in well-known autonomous driving systems.

Despite the relative success of DeepXplore, DeepTest, and DeepRoad, in increasing the test coverage of neural networks, Ma et al.~\cite{deepCT} remarked that the runtime state space is very large when each neuron output is considered as a state, which can lead to a combinatorial explosion. To help address this issue, they proposed DeepCT, which is a new testing method that adapts combinatorial testing (CT) techniques to deep learning models, in order to reduce the testing space. 
CT~\cite{combinatorial} has been successfully applied to test traditional software requiring many configurable parameters. It helps to sample test input parameters from a huge original space that are likely related to undetected errors in a program. For example, the $t$-way combinatorial test set covers all the interactions involving $t$ input parameters, in a way that expose efficiently the faults under the assumption of a proper input parameters' modeling. 
In DeepCT, K-way CT is adapted to allow for selecting effectively samples of neuron interactions inside different layers with the aim of decreasing the number of test cases. Given the initial test data sets. DeepCT generates some DNN-related K-way coverage criteria using constraint-based solvers (\ie{} by linear programming using the CPLEX solver~\cite{CPLEX}). Next, a new test data is generated by perturbing the original data within a prefixed value range, while ensuring that previously generated CT coverage criteria are satisfied on each layer. Ma et al.~\cite{deepCT} conducted an empirical study, comparing the 2-way CT cases with random testing in terms of the number of adversarial examples detected. They observed that random testing was ineffective even when a large number of tests were generated. In comparison, DeepCT performed well even when only the first several layers of the DNN were analyzed, which shows some usefulness for their proposed CT coverage criteria in the context of adversarial examples detection and local-robustness testing.\\
However, even though solvers like CPLEX represents the state-of-the-practice, their scalability remains an issue. Hence, the effectiveness of the proposed DeepCT approach on real-world problems using large and complex neural networks remains to be seen.

Sun et al.~\cite{sun2018testing} examined the effectiveness of the neuron coverage metric introduced by DeepXplore and report that a 100\% neuron coverage can be easily achieved by a few test data points while missing multiples incorrect behaviors of the model. To illustrate this fact, they showed how 25 randomly selected images from the MNIST test set yield a close to 100\% neuron coverage for an MNIST classifier. Thereby, they argue that testing DNNs should take into account the semantic relationships between neurons in adjacent layers in the sense that deeper layers use previous neurons' information represented by computed features and summarize them in more complex features. To propose a solution to this problem, they adapted the concept of Modified Condition/Decision Coverage (MC/DC)\cite{McDcCoverage} developed by NASA. The concepts of ``decision'' and ``condition'' in the context of DNN-based systems correspond to testing the effects of first extracted less complex features, which can be seen as potential factors, on more complex features which are intermediate decisions. Consequently, they specify each neuron in a given layer as a decision and its conditions are its connected input neurons from the previous layer.\\
They propose a testing approach that consists of a set of four criteria inspired by MC/DC and a test cases generator based on linear programming (LP). As an illustration of proposed criteria, we detail their notion of \emph{SS} coverage, which is very close to the spirit of MC/DC. Since the neurons' computed outputs are numeric continuous values, the \emph{SS} coverage could not catch all the interactions between neurons in successive layers. To address this limitation, three additional coverage criteria have been added to allow detecting different ways of how the changes of the conditions can affect the models' decision. 

Sun et al.~\cite{deepCover} also applied concolic testing~\cite{cute} to DNNs. Concolic testing combines concrete executions and symbolic analysis to explore the execution paths of a program that are hard to cover by blind test cases generation techniques such as random testing. The proposed adaptation of concolic testing to DNNs leverages state-of-the-art coverage criteria and search approaches. The authors first formulate an objective function that contains a set of existing DNN-related coverage requirements using Quantified Linear Arithmetic over Rationals (QLAR). Then, their proposed method incrementally finds inputs data that satisfy each test coverage requirement in the objective. Iterating over the set of requirements, the concolic testing algorithm finds the existing test input that is the closest data point to satisfy the current requirement following an evaluation based on concrete execution. Relying on this found instance, it applies symbolic execution to generate a new data point that satisfies the requirement and adds it to the test suite. The process finishes by providing a test suite that helps reaching a satisfactory level of coverage. To assess the effectiveness of their proposed approach, they evaluated the number of adversarial examples that could be detected by the produced test suites.


\subsubsection{Approaches that aim to detect errors in ML code implementations}
Given the stochastic nature of most ML algorithms and the absence of oracles, most existing testing techniques are inadequate for ML code implementations. As a consequence, the ML community have resorted to numerical testing, property-based testing, metamorphic testing, mutation testing, coverage-guided fuzzing testing, and proof-based testing techniques to detect issues in ML code implementations. In the following, we present the most prominent techniques. 

\paragraph{\textbf{Numerical-based testing: Finite-difference techniques}}
Most machine learning algorithms are formulated as optimization problems that can be solved using gradient-based optimizers, such as gradient descent or L-BFGS (\ie{} Limited-memory Broyden–Fletcher–Goldfarb–Shanno algorithm). The correctness of the objective function gradient that are computed with respect to the model parameters, is crucial. In practice, developers often check the accuracy of gradients using a finite difference technique that simply consists in performing the comparison between the analytic gradient and the numerical gradient. However, because of the increasing complexity of models' architectures, this technique is prone to errors. 
To help improve this situation, Karpathy~\cite{CS231n} have proposed a set of heuristics to help detect faulty gradients. 

\paragraph{\textbf{a)Use of the centered formula}}
Instead of relying on the traditional gradient formula, Karpathy recommends using the centered formula from Equation~\ref{centered_formula} which is more precise. 
The taylor expansion of the numerator indicates that the centered formula has an error in the order of $O\left(h^2\right)$, while the standard formula has an error of $O\left(h\right)$. 
\begin{equation}
\label{centered_formula}
\frac{df\left(x\right)}{dx}=\frac{f\left(x+h\right)-f\left(x-h\right)}{2h}
\end{equation}

\paragraph{\textbf{b) Use of relative error for the comparison}}
As mentioned above, developers perform gradient checking by computing the difference between the numerical gradient $f_n^\prime$ and the analytic gradient $f_a^\prime$. This difference can be seen as an absolute error and the aim of the gradient checking test is to ensure that it remains below a pre-defined fixed threshold. With deep neural networks, it can be hard to fix a common threshold in advance for the absolute error. Karpathy recommends fixing 
a threshold for relative errors. So, for a deep neural network's loss that is a composition of ten functions, a relative error \ref{relative_error} of $1\exp^{-2}$ might be acceptable because the errors build up through backpropagation. Conversely, an error of $1\exp^{-2}$ for a single differentiable function likely indicates an incorrect gradient.
\begin{equation}
\label{relative_error}
\frac{|f_a^\prime-f_n^\prime|}{max(|f_a^\prime|,|f_n^\prime|)}
\end{equation}

\paragraph{\textbf{c) Use of double precision floating point}}
Karpathy recommends to avoid using single precision floating point to perform gradient checks, because it often causes high relative errors even with a correct gradient implementation.

\paragraph{\textbf{d) Stick around active range of floating point}}
To train complex statistical models, one needs large amounts of data. So, it is common to opt for mini-batch stochastic gradient descent and to normalize the loss function over the batch. However, if the back-propagated gradient is very small, additional divisions by data inputs count will yield 
extremely smaller vales, which in turn can lead to numerical issues. As a solution to this issue, Karpathy recommends computing 
the difference between values with minimal magnitude, otherwise one should scale the loss function up by a constant to bring the loss to a denser floats range, ideally on the order of $1.0$ (where the float exponent is $0$).

\paragraph{\textbf{e) Use only few random data inputs}}
The use of fewer data inputs reduces the likelihood to cross kinks when performing the finite-difference approximation. These kinks refer to non-differentiable parts of an objective function and can cause inaccuracies in the numerical gradients. For example, let's consider an activation function \textit{ReLU} that has an analytic gradient at the zero point, \ie{} it is exactly zero. The numerical gradient can compute a non-zero gradient because it might cross over the kink and introduce a non-zero contribution. Karpathy strongly recommends to perform the gradient checking for a small sample of the data and infer the correctness of gradient for an entire batch, because it makes this sanity-check fast and more efficient in practice.

\paragraph{\textbf{f) Check only few dimensions}}
Recent statistical models are more and more complex and may contain thousands parameter with millions of dimensions. The gradients are also multi-dimensional. To mitigate errors, Karpathy recommends checking a random sample of the dimensions of the gradient for each separate model's parameter, while assuming the correctness of the remaining ones. 

\paragraph{\textbf{g) Turn off dropout/regularization penalty}}
Developers should be aware of the risk that the regularization loss overwhelms the data loss and masks the fact that there exists a large error between the analytic gradient of data loss function and its numerical one. Such circumstance can result in a failure to detect 
an incorrect implementation of the loss function or its corresponding gradient using the finite-difference approximation technique. Karpathy recommends turning off regularization and checking the data loss alone, and then the regularization term, independently. 
Moreover, recent regularization techniques applied to deep neural networks such as dropout, induce a non-deterministic effect when performing gradient check. Thereby, to avoid errors when estimating the numerical gradient, it is recommended to turn them off. An alternative consists in forcing a particular random seed when evaluating both gradients. 

\paragraph{\textbf{Property-based testing}} Property-based testing is a technique that consists in inferring the properties of a computation using the theory and formulating invariants that should be satisfied by the code. Using the formulated invariants, test cases are generated and executed repeatedly throughout the computation to detect potential errors in the code.
Using property-based testing, one can ensure that probability laws hold throughout the execution of a model. For example, one can test that all the computed probability values are non-negatives. For a discrete probability distribution, as in the case of a classifier, one can verify that the probabilities of all events add up to one. Also, marginalization can be applied to test probabilistic classifiers.\\
Roger et al.~\cite{MCMCTesting} applied property-based testing to detect errors in implementations of MCMC (Markov chain Monte Carlo) samplers. They derived by hand the update rules for individual variables using the theory, \ie{} they sample a random variable from its conditional distribution. Then, they wrote a single iterative routine including those rules. To test the correctness of the produced samples, they verified that the conditional distribution was consistent with the joint distribution at each update iteration.

Karpathy~\cite{CS231n} recommend the verification of the following properties. 
\begin{description}
\item[Initial random loss :] when training a 
neural network classifier, 
turn off the regularization by setting its corresponding strength hyperparameter to zero and verify that initial softmax loss is equal $-\log\left(\frac{1}{N_c}\right)$ with $N_c$: number of label classes. One expect a diffuse probability of $\frac{1}{N_c}$ for each class.
\item[Overfitting a tiny dataset :] Keeping the regularization term turned off, extract a sample portion of data, (one or two examples inputs from each class) in order to ensure that the training algorithm can achieve efficiently zero loss. 
    Breck et al. also recommend watching the internal state of the model on small amounts of data with the aim of detecting issues like numerical instability that can induce invalid numeric values like NaNs or infinities. 
\item[Regularization role :] increase the regularization strength and check if the data loss is also increasing.
\end{description}

Another testing technique that shares the same philosophy as property-based testing is metamorphic testing.

\paragraph{\textbf{Metamorphic testing}}
Murphy et al.~\cite{murphyProps} introduced metamorphic testing to ML in 2008. They defined several Metamorphic Relationships (MRs) that can be classified into six categories (\ie{} additive, multiplicative, permutative, invertive, inclusive, and exclusive). They applied these MRs to test three ML applications: Marti-Rank, SVM-Light (support vector machine with a linear kernel), and PAYL~\cite{murphyMT}. Xie et al.~\cite{xieMT} expanded the work of Murphy et al. by introducing new MRs for testing K-Nearest Neighbors classifier and Naive Bayesian classifier.\\
Recent research works~\cite{LastMT} have investigated the application of metamorphic testing to more complex machine learning algorithms such as SVM with non-linear kernel and deep residual neural networks (ResNET). The technique was able to successfully find mutants in open-source machine learning applications.

\paragraph{\textbf{Mutation testing}}
Ma et al.\cite{deepmutation} proposed DeepMutation that adapts mutation testing~\cite{mutation} to DNN-based systems with the aim of evaluating the test data quality in terms of its capacity to detect faults in the programs. 
Mutation testing consists in injecting artificial faults (\ie{} mutants) in a program under test and generating test cases to detect them.
To build DeepMutation, Ma et al. defined a set of source-level mutation operators to mutate the source of a ML program by injecting faults. These operators allow injecting faults in the training data (using data mutation operators) and the model training source code (using program mutation operators). After the faults are injected, the ML program under test is executed, using the mutated training data or code, to produce the resulting mutated DNNs. The data mutation operators are intended to introduce potential data-related faults that could occur during data engineering (\ie{} during data collection, data cleaning, and--or data transformation). Program mutation operators' mimic implementation faults that could potentially exist in the model implementation code. These mutation operators are sematic-based and specialized for DNNs' code.
Training models from scratch following source-level mutations is very time-consuming since advanced deep learning systems require often tens of hours and even days to learn the model's parameters. Moreover, manually designing specific mutation operators using information about faults occurring in real-world DNNs systems is challenging, since it is difficult to imagine and simulate all possible faults that occur in real-world DNNs. 
To circumvent the cost of multiple re-execution and fill in the gap between real-world erroneous models and mutated models, the authors define model-level mutation operators to complement source-level mutation operators. These operators directly change the structure and the parameters of neural network models to scale the number of resulted mutated models for testing ML programs in an effective way, and for covering more fine-grained model-level problems that might be missed by only mutating training data and--or programs.
Once the mutated models are produced, the mutation testing framework assesses the effectiveness of test data and specify its weaknesses based on evaluation metrics related to the killed mutated models count. ML engineers can leverage this technique to improve data generation and increase the identification of corner-cases DNN behaviors.

\paragraph{\textbf{Coverage-Guided Fuzzing}}
Odena and Goodfellow~\cite{tensorfuzz} developed a coverage-guided fuzzing framework specialized for testing neural networks. Coverage-guided fuzzing has been used in traditional software testing to find critical errors. For ML code, the fuzzing process consists of handling an input corpus that evolves through the execution of tests by applying random mutation operations on its contained data and keeping only interesting instances that allow triggering new program behavior. 
Iteratively, the framework samples an input instance from testing data corpus and mutates it in a way that preserves the correctness of its associated label. Then, it checks whether the corresponding state vector is meaningfully different from the previous ones using a fast approximate nearest neighbor algorithm based on a pre-specified distance. Each input data that is relatively far from the existing nearest neighbor, is added to the test cases set.
The framework, entitled TensorFuzz, is implemented to test TensorFlow-based neural network models automatically. The effectiveness of the proposed testing approach was assessed using three known issues in neural networks' implementations. Results show that TensorFuzz surpasses random search in : (1) finding NaNs values in neural network models trained using numerical instable cross-entropy loss, (2) generating divergences in decision between the original model that is encoded with 32-bit floating point real and its quantized version that is encoded with only 16-bit, and (3) surfacing undesirable behavior in character level language RNN models.

\paragraph{\textbf{Proof-based testing}}
Selsam et al.\cite{bugfreeMLS} proposed to formally specify the computations of ML programs and construct formal proofs of written theorems that define what it means for the programs to be correct and error-free. Using the formal mathematical specification of a ML program and a machine-checkable proof of correctness representing a formal certificate that can be verified by a stand-alone machine without human intervention, ML developers are able to find errors in the code.\\
Using their proposed approach, Selsam et al. analyzed a ML program designed for optimizing over stochastic computation graphs, using the interactive proof assistant \textit{Lean}, and reported it to be bug-free. However, although this approach allows to detect errors in the mathematical formulation of the computations, it cannot help detect execution errors due to numerical instabilities, such as the replacement of real numbers by floating-point with limited precision.\\

\section{Conclusion and future research}\label{sec:conclusion}
The inductive nature of ML programs makes it difficult to reason about their behavior. Recently, researchers have started to develop new testing techniques to help ML developers detect debug and test ML programs. In this paper, we describe the generic process of a ML program creation, from data preparation to the deployment of the model in production, and explain the main sources of faults in a ML program. Next, we review testing techniques proposed in the literature to help detect these faults both at the model and implementation levels; explaining the context in which they can be applied as well as their expected outcome. We also identify gaps in the literature related to the testing of ML programs and suggest future research directions for the scientific community. By consolidating the progress made in testing ML programs in a single document, we hope to equip ML and software engineering communities with a reference document on which future research efforts can be built. Practitioners can also use this paper to learn about existing testing techniques for ML programs, which in turn can help improve the quality of their ML programs. %

\balance
\bibliography{OnTestingMLApps}

\end{document}